# Two-dimensional octagon-structure monolayer of nitrogen group elements and the related nano-structures

Yu Zhang[a], Jason Lee[a], Wei-Liang Wang[a], Dao-Xin Yao[a,b*]

[a] State Key Laboratory of Optoelectronic Materials and Technologies, School of Physics and Engineering, Sun Yat-Sen University, Guangzhou 510275, China

[b] Beijing Computational Science Research Center, Beijing 100084, China

[*] Corresponding author: yaodaox@mail.sysu.edu.cn

ABSTRACT

In the purpose of expanding the family of two-dimensional materials, we predict the existence of two-dimensional octa-structure of nitrogen group elements that are composed of squares and octagons in first-principle method based on density functional theory (DFT). From our calculations, electronic structures of all monolayers show that they are semiconductors with indirect (N, P, Bi) and direct (As, Sb) band gaps (0.57-2.61eV). Nano-ribbons of three different unpassivated edges and their band structures are also investigated. Because of the reconstruction on the edges and dangling bonds, there exist ferromagnetic edge states in P, As, Sb nano-ribbons with different edges, and a Dirac point near π is found in the band structure of one specific N nano-ribbon. These structures may be useful in future applications, such as semiconductor devices, spintronics, hydrogen storage and quantum computation.

## 1. Introduction

Since graphene was successfully exfoliated from graphite in 2004[1], numerous theoretical and experimental researches have been invested in studying its properties[2-5]. Graphene possesses exceptional properties in many aspects, especially for its electrical property on account of the existence of Dirac electrons. Triggered by these studies, a great number of graphene derivatives were proposed and investigated in the past years, such as graphane[6,7], graphyne[8,9], graphdiyne[10], octagraphene[11], graphenylene[12,13], and biphenylene[14]. Besides carbon, many other elements can exist in two-dimensional (2D) monolayers and exploring more 2D materials has great importance for various applications.

Phosphorus has many allotropes with different structures. It is worth mentioning that a layered allotrope of phosphorus, black phosphorus[15-17], which has been studied decades ago, is a semiconductor with a nonzero band gap and high mobility. These properties make black phosphorus a better choice for applications[18] compared to other 2D materials, and it can be fabricated into field-effect transistors with excellent performance in room temperature[19,20]. In 2014, a new layered structure of phosphorus, which has buckled honeycomb structure similar to silicene, is investigated theoretically and named blue phosphorus[21-23]. It exhibits a much larger band gap than black phosphorus. In addition, honeycomb monolayers of other nitrogen group (VA) elements (N, P, As, Sb, Bi) are also predicted to exist[24].

Octagon and square structures can exist in carbon monolayers[11-14]. In this work, we try to explore a new 2D allotrope of VA elements with octagons and squares in their structures, as shown in Fig.1. There are five valence electrons in one atom of VA elements, and each atom is bonded to three neighbors similar to the honeycomb structure, so we expect them to have similar structure parameters and properties.

First-principle method based on density functional theory (DFT) is used to investigate both monolayers and three different nano-ribbons with unpassivated edges. We perform structure optimization to obtain stable structures, and calculate the electronic structures to explore their electronic and magnetic properties. 2D monolayers turn out to be semiconductors with large band gaps, and various reconstructions happen at the edge of nano-ribbons, which finally result in distinctive properties. These properties can be used in future applications.

## 2. Computation methods and models

### 2.1. Computational methods

In this paper, new octagon phases of VA elements are studied in first-principle methods. Our calculation is based on Plane Augmented Wave (PAW)[25,26] with Perdew-Burke-Ernzerh (PBE) of exchange-correlation as implemented in the Vienna Ab initio Simulation Package (VASP) code[27]. We use default plane-wave cutoff energy during the calculation. The Monkhorst-Pack[28] scheme for k-point samplings with 15×15×1 mesh for planar sheets and 5×1×1 mesh for nano-ribbons were used to sample the Brillouin zone. A vacuum of about 15 Å thick is inserted between two layers and two nanoribbons to avoid interactions. The structures are optimized until the remnant Hellmann-Feynman forces on the ions are less than 0.01 eV/Å. Spin-orbit Coupling (SOC) is included only in the calculations of band structures of monolayers.

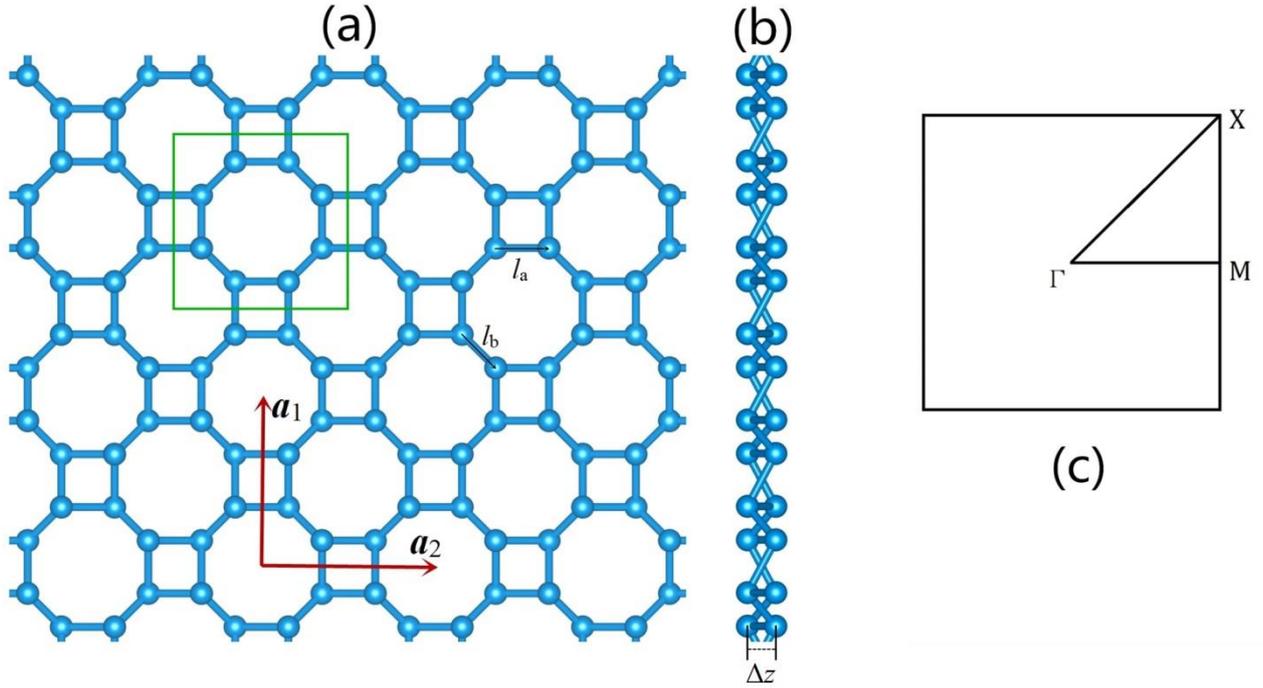

Fig. 1 (a) Top view and (b) side view of the investigated monolayer structure. The green square presents the unit cell, and the red arrows $a_1$, $a_2$ show two basis vectors of the unit cell. There are 8 atoms in each unit cell. $l_a$ and $l_b$ are two type bonds, and $\Delta z$ is the buckled constant. (c) The first Brillouin zone.

*2.2. Computational models*

We perform structure optimizations to obtain the stable lattice structures shown in Fig. 1 (a) and (b). From the top view (as shown in Fig. 1 (a)), the structures consist of squares and octagons instead of hexagons in their honeycomb structures. The unit cell is a square, and the two basis vectors are given by $a_1=ai$, $a_2=aj$, where $a$ is the lattice constant. While from the side view (as shown in Fig. 1 (b)), we can see squares composed of interlace buckled zigzags. There are 8 atoms per unit cell rather than 4, that is because corresponding atoms in the adjacent squares are not in the same plane. The buckled octa-structure presents $D_4$ symmetry different from the honeycomb structure. We calculate the bands of monolayers along $\Gamma(0,0)$-M$(0,\pi)$-X$(\pi,\pi)$-$\Gamma(0,0)$ in the first Brillouin zone shown in Fig. 1 (c).

Table 1 Structural parameters (unit in Å) of monolayers

| Element | $a$ | $\Delta z$ | $l_a$ | $l_b$ |
|---|---|---|---|---|
| N | 4.4187 | 0.6972 | 1.5173 | 1.4031 |
| P | 6.4564 | 1.2481 | 2.2723 | 2.2567 |
| As | 7.1020 | 1.4196 | 2.5267 | 2.5070 |
| Sb | 8.1326 | 1.6573 | 2.9088 | 2.8917 |
| Bi | 8.5972 | 1.7358 | 3.0606 | 3.0552 |

Optimized structural parameters of N, P, As, Sb, Bi are listed in Table.1. $a$ is the lattice constant, $\Delta z$ is the buckled constant as shown in Fig. 1 (b), $l_a$ is the length of bond in the square, and $l_b$ is the length of bond connecting two squares, as shown in Fig. 1 (a). We can tell that lattice constant $a$, buckled constant $\Delta z$ and bond length $l_a$ and $l_b$ are all monotonously increasing with respect to increasing atomic number from N to Bi, and are close to the results of the honeycomb structures[24]. In addition, for all elements, bond length $l_a$ is larger than $l_b$. Comparing with parameters of the hexagonal structures, bond length $l_a$ is a bit longer and $l_b$ is slightly shorter.

As for the stability of monolayers, geometry optimizations present an energetically stable configuration for all the materials. For purpose of proving the stability further, vibration frequencies at $\Gamma$ are calculated to ascertain whether they are dynamically stable, and there is no imaginary vibration mode. In order to compare the stability of octagon monolayers and hexagon ones, we calculate their cohesive energies defined as $E_{coh} = E_{monolayer} - E_{isolated}$, which means the energy difference between the energy of one isolated atom and the energy of one atom in the monolayers. $E_{coh}$ for octagon monolayers are -6.45eV (N), -5.20eV (P), -4.50 (As), -3.91eV (As), -3.64eV (Bi). The cohesive energies per atom for octagon monolayers are slightly higher than that of the honeycomb counterparts, and the energy differences are 0.36eV (N), 0.16eV (P), 0.14eV (As), 0.13eV (Sb), 0.12eV (Bi).

## 3. Results and discussion

### 3.1 Electronic Structure of Monolayers

Using DFT method, we calculate the electronic structures of the five VA elements with this special structure, as shown in Fig.2. They are all semiconductors, and the band gaps from the calculations are listed in table.2. By comparing band structures with ((f)-(j)) and without ((a)-(d)) SOC, we can see that SOC has no effect on band structures of N and P. For As, Sb and Bi, SOC breaks energy degeneracy of bands from X to K. In addition, the effect of SOC increases as the atomic number of elements increases.

As shown in Fig. 2, band gap decreases monotonously from 2.61 to 0.57eV (0.26eV) without (with) SOC. Without SOC, monolayers of N and P have indirect gaps and monolayers of As, Sb and have direct gaps.The effect of SOC of N and P monolayers is so small that we find no changes in our calculation. But for As, Sb, Bi, the band gaps decrease obviously by including SOC. Furthermore, SOC changes the gap of Bi monolayer from direct gap to an indirect one.

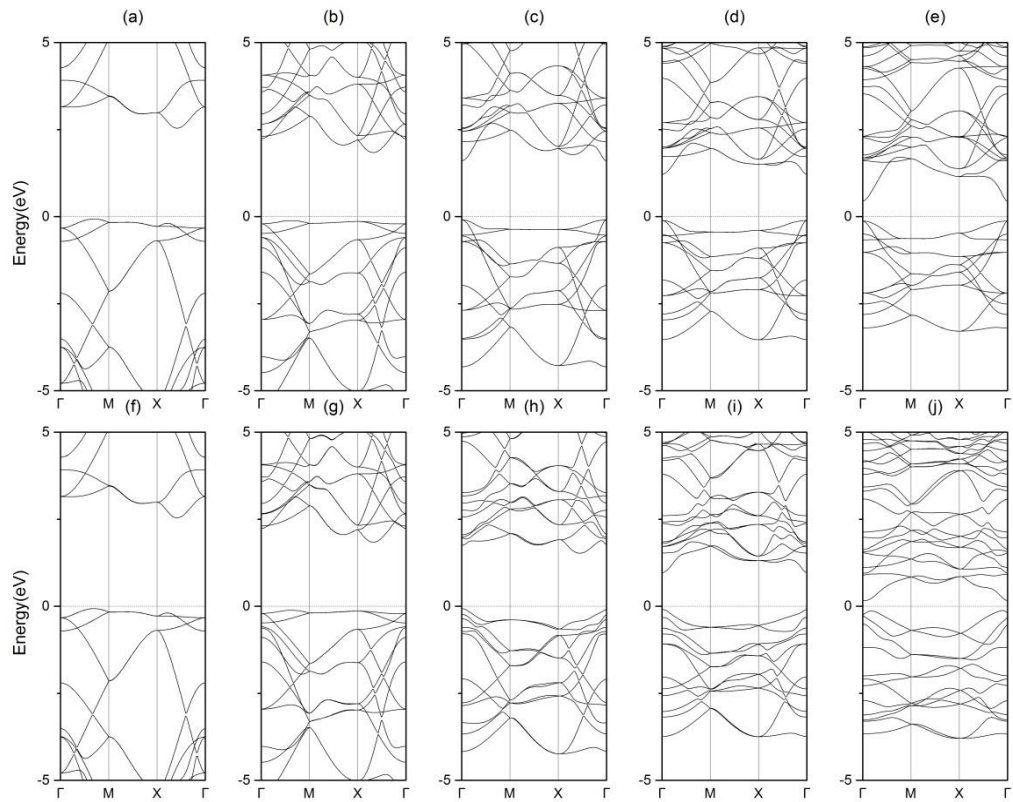

Fig. 2 Band structures of (a)N, (b)P, (c)As, (d)Sb and (e)Bi without SOC and (f)N, (g)P, (h)As, (i)Sb, (j)Bi with SOC.

Table 2 Table 2 Band gaps (unit in eV) of monolayer of VA elements

| Element | Without SOC | Gap type | With SOC | Gap type | Change of band gap |
|---------|-------------|----------|----------|----------|--------------------|
| N | 2.61 | Indirect | 2.61 | Indirect | 0 |
| P | 1.95 | Indirect | 1.95 | Indirect | 0 |
| As | 1.70 | Direct | 1.59 | Direct | 0.11 |
| Sb | 1.33 | Direct | 1.05 | Direct | 0.28 |
| Bi | 0.57 | Direct | 0.26 | Indirect | 0.31 |

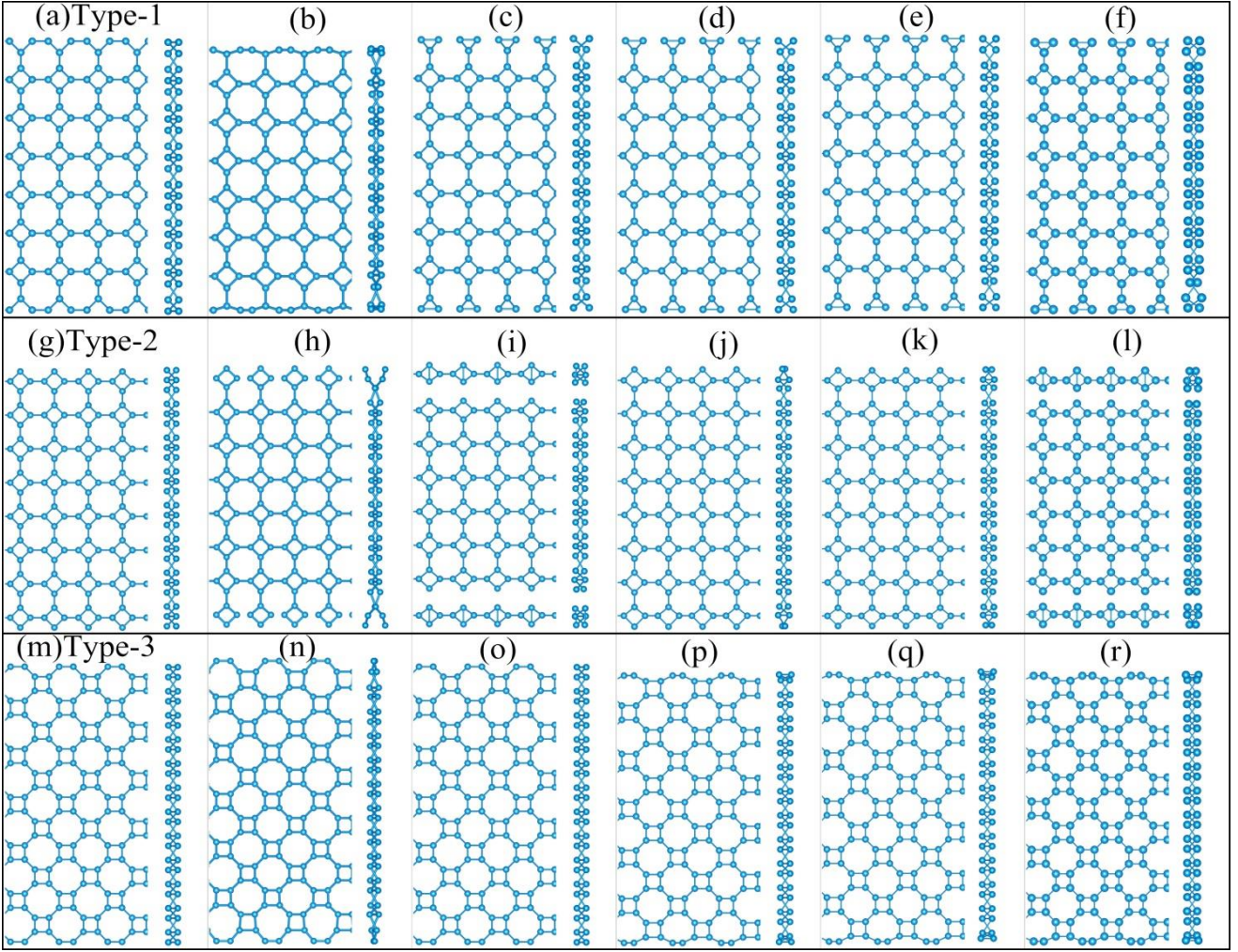

Fig. 3 1D Structure of each element with (a)type-1, (g)type-2 and (m)type-3 edge before optimization; optimized structure of (b)N, (c)P, (d)As, (e)Sb and (f)Bi with type-1 edge; optimized structure of (h)N, (i)P, (j)As, (k)Sb and (l)Bi with type-2 edge; optimized structure of (n)N, (o)P, (p)As, (q)Sb and (r)Bi with type-1 edge.

## 2.2 Three Different Nano-ribbons

We cut the monolayer to nano-ribbons with three different edges from different directions and locations, as shown in Fig.3 (a), (g) and (m), and they are defined as type-1, type-2, and type-3. It needs to be stressed that the edges are not passivated in our calculations.

In this paper, we consider the three different nano-ribbons with 22, 24, and 24 atoms wide, respectively. Before the calculation of electronic structures, no less than four outermost atoms of each side were relaxed during the structure optimizations. Reconstruction happens in many cases, as shown in Fig.3 (b)-(f), (h)-(l), (n)-(r), and Bi (i), P (l) with type-2 edge may not stable since bonds are broken after structure optimizations. It is interesting to notice that the edge of N type-2 nano-ribbon (Fig. 3 (h)) reconstructs to the structure with two adjacent squares stretch up and down. Because of the reconstruction of atoms at the edges, the band structures turn out to be Intriguing.

Fig.4 shows the band structures of nano-ribbons with three different edges. Spin-up bands and spin-down bands are degenerate, except for As, Sb with type-2 edge and P with type-3 edge. We notice that there are eight bands near the Fermi level, the four bands above Fermi level and the four bands below the Fermi level belong to different spin components. From our calculations, magnetic moments are about 4μB in the cases of As(3.98$\mu_B$), Sb(4.16$\mu_B$) with type-2 edge and P(4.03$\mu_B$) with type-3 edge. We analyze the charge density distributions of these eight bands, and find that the charge of these bands is mainly distributed on the edges of the nano-ribbon, as shown in Fig.5. We can conclude that these eight bands are corresponding to eight edge states which may be cause by dangling bonds.

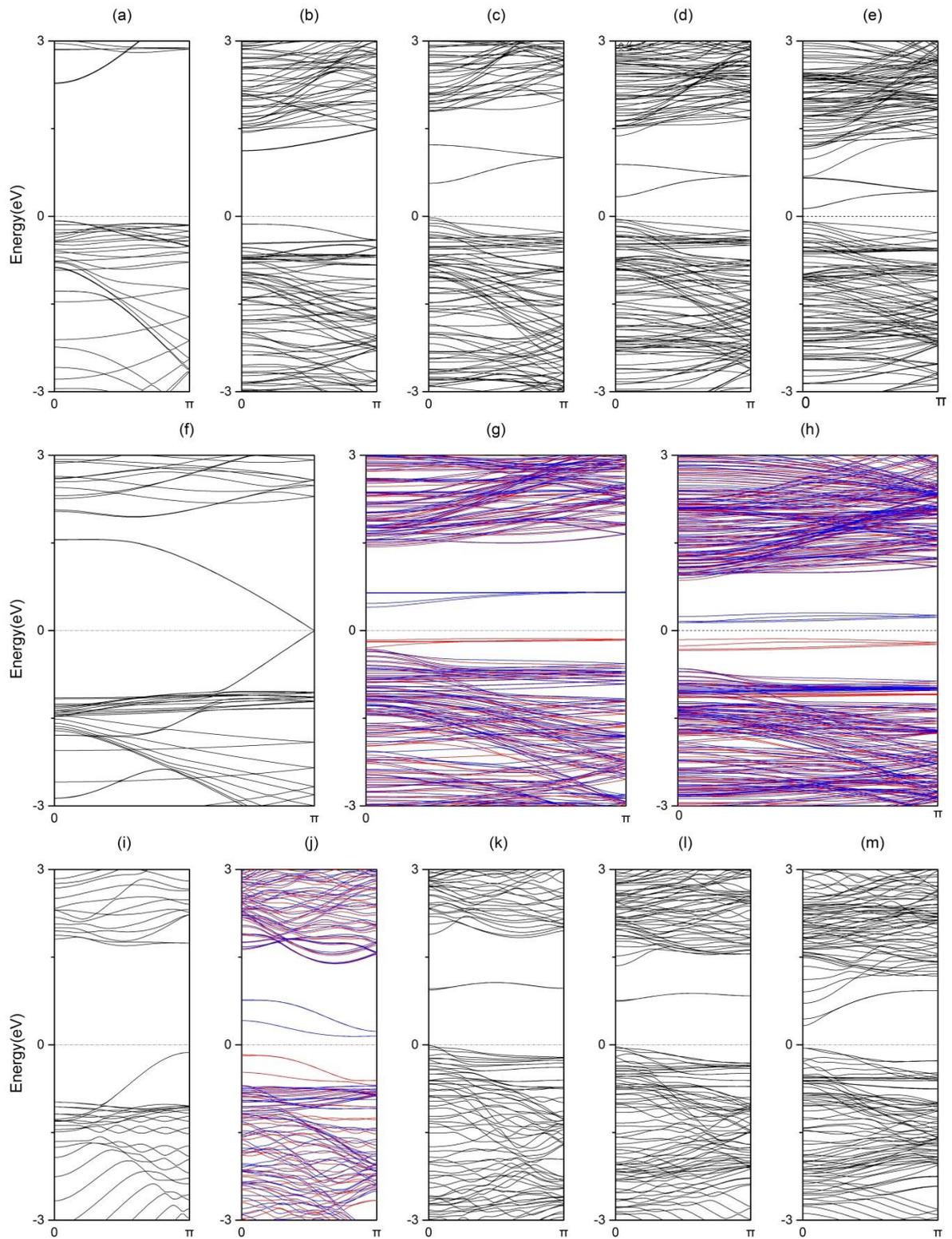

Fig. 4 Band structures of nano-ribbons of (a)N, (b)P, (c)As, (d)Sb and (e)Bi with type-1 edge, (f)N, (g)As and (h)Sb with type-2 edge, and (i)N, (j)P, (k)As, (l)Sb and (m)Bi with type-3 edge. Bands without spin polarization are plot in black lines. For (g), (h) and (j), black and red lines represent spin-up and spin-down bands, respectively.

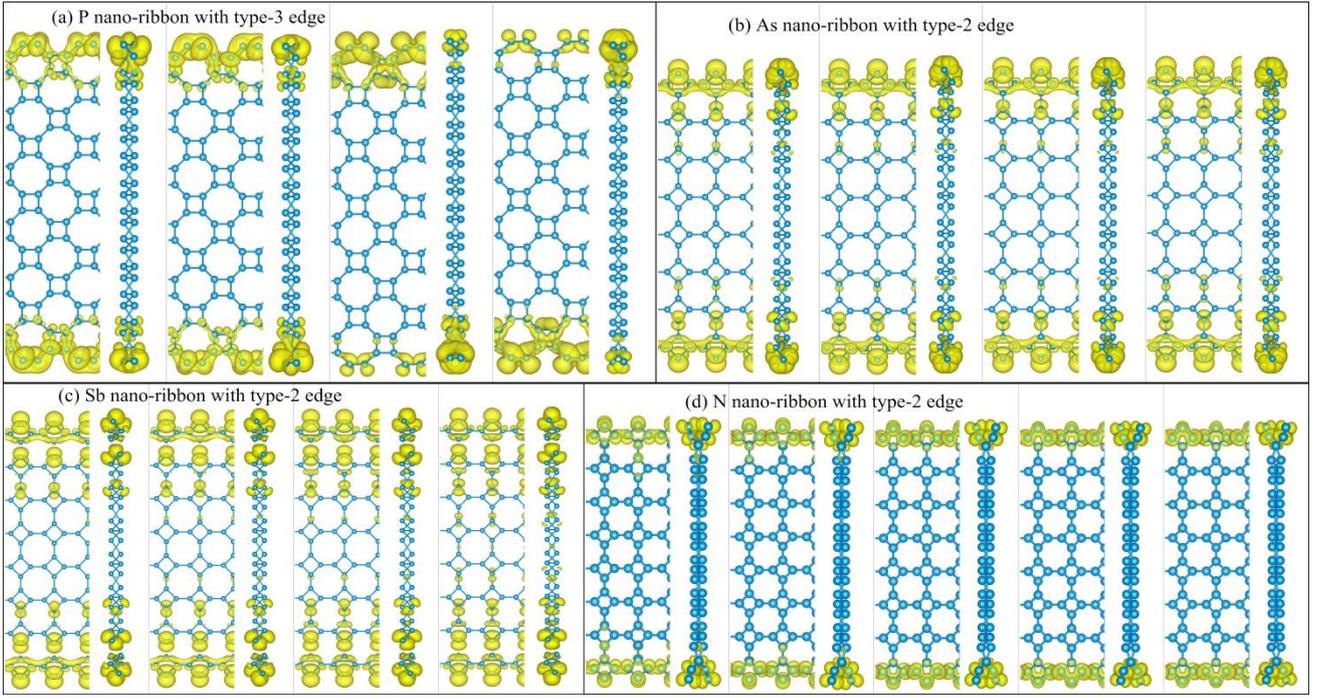

Fig. 5 Charge density distributions of the four bands under Fermi level of (a)P nano-ribbon with type-3 edge,(b) As nano-ribbon with type-2 and (c) Sb nano-ribbon with type-2 edge edge. (d) Charge density distributions of the band above Fermi level of N nano-ribbon with type-2 edge, at 0, 0.2π, 0.4π, 0.6π, 0.8π and π, respectively.

It is interesting to notice that two linear bands seem crossing at π in the first Brillouin Zone for N with type-2 edge, which might indicate a Dirac point existing. To make further efforts, we calculate the band structure near π(0.98π~π), which reveals more exquisite image, as shown in Fig. 6 (a). When SOC is not included, Fermi level exactly crosses the Dirac point at 0.9976π rather than π with zero gap. This special character is caused by the extraordinary rearrangement at the edge of the ribbon. We analyze the charge density of the band below Fermi level at 0, 0.2π, 0.4π, 0.6π, 0.8π and π of the reciprocal vector to find that edge effect subducts from the edge to inside to a extent of about 0.7nm, as shown in Fig.5 (d). When SOC is included the Dirac point move to 0.9996π, as shown in Fig.6 (b).

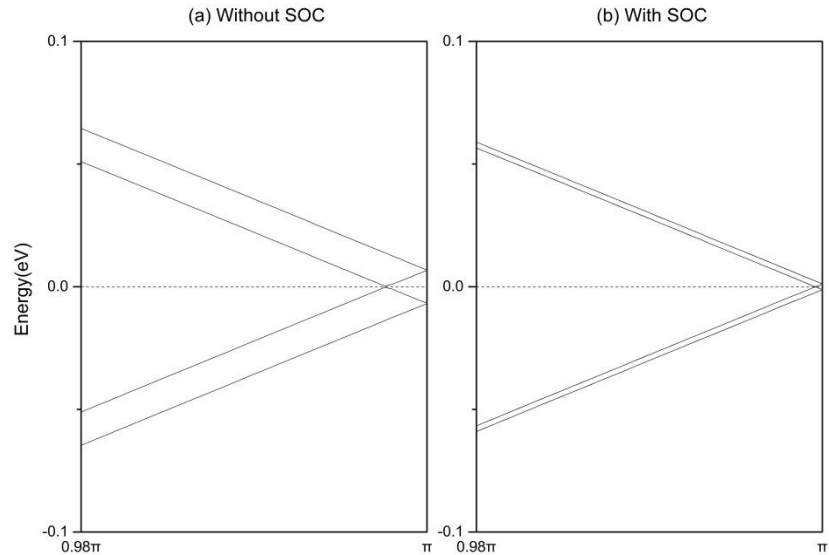

Fig. 6 Band structures of N type-2 nano-ribbon (a) without SOC, (b) with SOC

## 4. Conclusions

We obtain the stable structures of the octagon phases of VA elements based on first-principle calculation. 2D electronic structures show that they are semiconductors with band gaps that decrease monotonously from N to Bi. SOC, which is more important for heavier element, breaks energy degeneracy of bands.

By studying the nano-ribbons, we get various optimized edge reconstruction, and ferromagnetic edge states exist in some cases with four spin-polarized bands lying both below and above Fermi level. The rearrangement of nano-ribbon of N with type-2 edge results in edge states with a Dirac point near point π in Brillouin zone. These distinct electronic properties may be cause by the dangling bonds at the edges.

Due to their intrinsic band gaps, 2D layers with octagon-structure can be potential candidates in the semiconductor devices, spintronics and quantum computation. Besides, many 2D materials are studied in the aspect of hydrogen storage. Hydrogen is adsorbed by metal atoms attached to the surface of the materials. According to the previous calculations[11], the hydrogen storage capacity of octagraphene absorbed with Ti is 7.76 wt%, which is close to that of graphene. Another 2D carbon allotrope, biphenylene, can have a stronger hydrogen storage capacity (11.6 wt.%) than graphene when Li is doped on its octagonal rings[29]. We expect that the monolayers of VA elements with octagon-structure can be a good candidate for hydrogen storage.

## Acknowledgements

The authors thank Wei-Feng Tsai, Fan Yang, and Feng-Chuan Chuang and for helpful discussions. This project is supported by National Basic Research Program of China (2012CB821400), NSFC-11275279, NSFC-11074310, Fundamental Research Funds for the Central Universities of China, National Supercomputer Center in Guangzhou, RFDPHE-20110171110026, and NCET-11-0547.

## Appendix

In order to explore the orbital feature of bands, we calculate the projections of each band to s, $p_x$, $p_y$ and $p_z$ orbitals, as shown in Fig. 7. Electrons in s orbital occupy the lowest eight bands without hybridizations to other orbitals except for N, and we plot the lowest eight bands in red lines to represent s orbital. Above those s orbital bands, $p_x$, $p_y$, and $p_z$ orbitals are dominant, which are represented by three different colors. For P, As, Sb and Bi, there is an energy gap between the top s band and the bottom p band, and the gap decreases from Bi to P, monotonously. In the case of N, two lowest band branches overlap with each other, which lead to the hybridization of s orbitals and p orbitals. We plot the lowest eight bands in red for N (Fig. 7 (a)). It is worth mentioning that $p_y$ orbital has the main contribution to the bands near the Fermi energy, and the energies of $p_x$ and $p_z$ bands are lower than $p_y$ band.

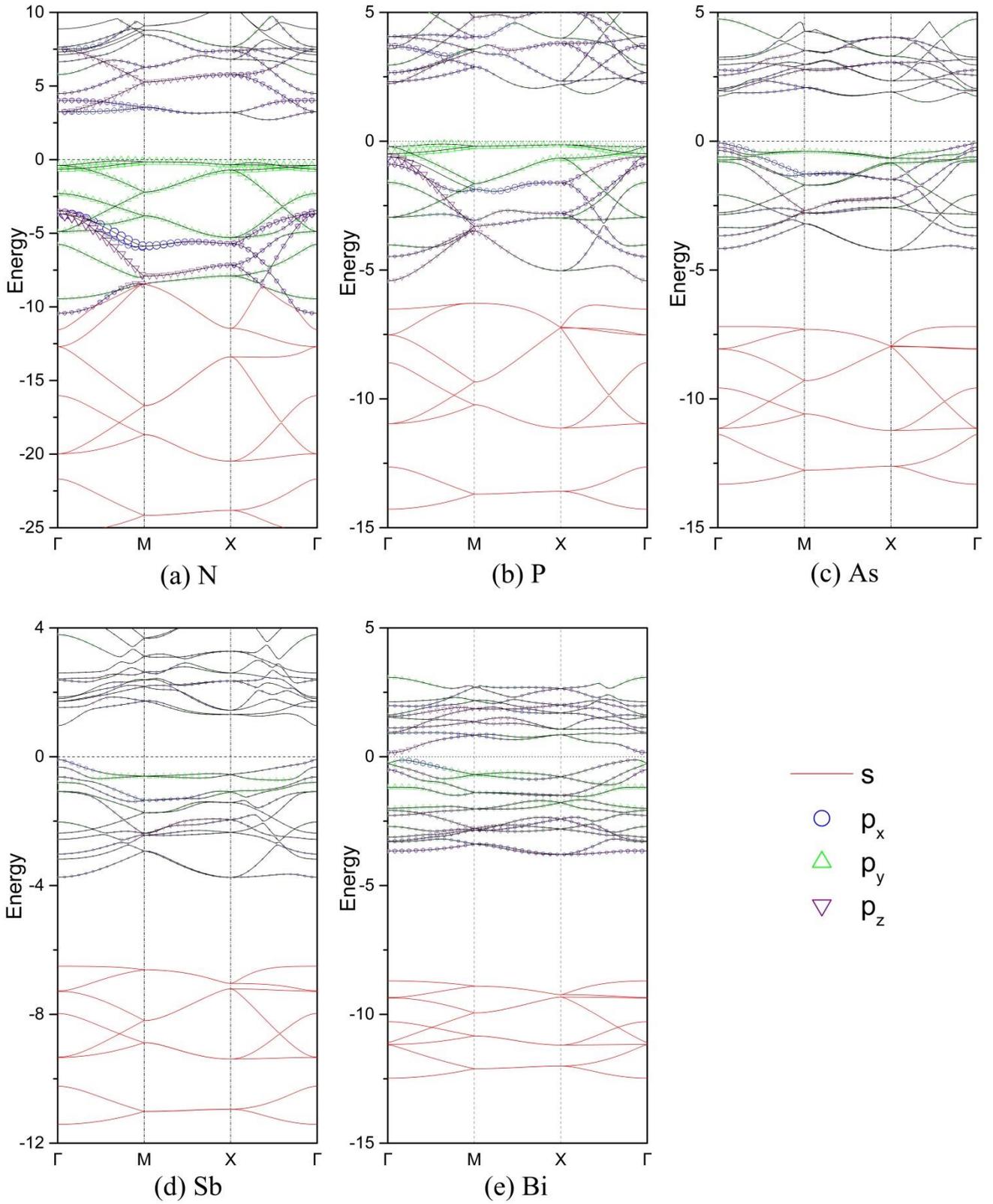

Fig. 7  Band projections of (a)N, (b)P, (c)As, (d)Sb and (e)Bi monolayer to s, $p_x$, $p_y$ and $p_z$ orbitals. Red lines correspond to s orbitals, three p orbitals are presented in three symbols.